\documentclass[pra,onecolumn,floatfix,superscriptaddress,longbibliography,notitlepage, nofootinbib]{revtex4-1}
\pdfoutput=1

\usepackage[utf8]{inputenc}
\usepackage{braket}
\usepackage{amsmath}

\usepackage{dcolumn}   
\usepackage{bm}        
\usepackage{amssymb}
\usepackage{mathtools}
\usepackage{tikz}
\usepackage{qcircuit}
\usepackage{appendix}
\usepackage{hyperref}
\usepackage{subcaption}
\usepackage{amsmath,amsfonts,amssymb}
\usepackage{mathtools}
\usepackage{thmtools,thm-restate}
\usepackage{algorithm}
\usepackage{mathrsfs}
\usepackage{tikz}
\usepackage{subcaption}
\usepackage{pgfplots}
\usetikzlibrary{3d}
\usepackage{tkz-euclide}

\usepackage{algpseudocode}

\newcommand{\Ibb}{\mathbb{I}}

\newtheorem{definition}{Definition}
\newtheorem{theorem}{Theorem}
\newtheorem{lemma}{Lemma}

\usetikzlibrary{arrows.meta}

\def\BibTeX{{\rm B\kern-.05em{\sc i\kern-.025em b}\kern-.08em
    T\kern-.1667em\lower.7ex\hbox{E}\kern-.125emX}}

\begin{document}
\title{Simulating Time-dependent Hamiltonian Based On High Order Runge-Kutta and Forward Euler Method}
\author{Nhat A. Nghiem}
\email{nhatanh.nghiemvu@stonybrook.edu}
\affiliation{Department of Physics and Astronomy, State University of New York at Stony Brook, Stony Brook, NY 11794-3800, USA}
\affiliation{C. N. Yang Institute for Theoretical Physics, State University of New York at Stony Brook, Stony Brook, NY 11794-3840, USA}

\begin{abstract}
We propose a new method for simulating certain type of time-dependent Hamiltonian $H(t) = \sum_{i=1}^m \gamma_i(t) H_i$ where $\gamma_i(t)$ (and its higher order derivatives) is bounded, computable function of time $t$, and each $H_i$ is time-independent, and could be efficiently simulated. Our quantum algorithms are based on high-order Runge-Kutta method and forward Euler method, where the time interval is divided into subintervals. Then in an iterative manner, the evolution operator at given time step is built upon the evolution operator at previous time step, utilizing algorithmic operations from the recently introduced quantum singular value transformation framework. 
\end{abstract}
\maketitle

\section{Introduction}
\label{sec: introduction}
Quantum simulation is one of the biggest potential application of quantum computer \cite{feynman2018simulating}. Beginning with a few pioneering works \cite{lloyd1996universal, aharonov2003adiabatic}, many quantum simulation algorithms have been proposed and improved. Notable works in time-independent regime includes \cite{berry2015hamiltonian, berry2007efficient, berry2012black, berry2009black, wiebe2011simulating, childs2012hamiltonian, low2017optimal}. Extension to time-dependent regime involves \cite{poulin2011quantum, berry2020time, chen2021quantum, an2022time, low2018hamiltonian, kieferova2019simulating}. The main obstacle in the simulation task is to translate the desired operator, e.g., evolution operator, into executable operations that could be implemented by a universal quantum computer. Thus, a metric for measuring the hardness of corresponding simulation is the number of elementary operations, or gates, required to construct the desired operator. A somewhat more practical metric is the depth of quantum circuit for implementing the evolution operator. 

The evolution of a quantum system under the Hamiltonian $H$ is governed by Schrodinger's equation, which is a first order homogeneous ordinary differential equation. In the time-independent regime, where the Hamiltonian of interest doesn't depend on time explicitly, most previous works explore the structure of given Hamiltonian, such as local interaction \cite{lloyd1996universal}, sparse-access to entries \cite{berry2007efficient, berry2015simulating, childs2012hamiltonian, low2017optimal}, linear combination of unitaries \cite{berry2015hamiltonian, childs2012hamiltonian}, to approximate the time evolution operator $\exp(-i H t)$. Optimal results have been found, showing that in order to simulate the evolution operator $\exp(-i H t)$ up to error $\epsilon$, a quantum circuit of depth $\mathcal{O}( d ||H|| t + \log(\frac{1}{\epsilon}))$ is required, where $d$ is the sparsity of $H$ and $||H||$ refers to the spectral norm of $H$. In the time-dependent regime, where Hamiltonian $H$ depends on time, methods based on Dyson series were ultilized in previous works \cite{poulin2011quantum, berry2020time, chen2021quantum, an2022time, low2018hamiltonian, kieferova2019simulating} to decompose the then evolution operator, which is $\exp(-i \int_{0}^t H(s)ds )$, into products of terms that could be handled. The key idea is that the integral $\int_0^t H(s) ds$ is approximated by a discrete summation, and its exponential could be further decomposed using Suzuki-Trotter formula. 

Most of the aforementioned works rely on an oracle that could compute the (time-dependent) entries of $H(t)$ at a specific time $t$. Here, we take an alternative route, inspired by ordinary differential equation numerical solver, e.g., Runge-Kutta and forward Euler method. We observe that the Schrodinger's equation itself is a first-order differential equation, and hence, the solution to such equation could be estimated via an iterative procedure in a similar manner to the popular ordinary differential equation solver, which is forward Euler method. We achieve our goal of simulating evolution operator in a given time interval by first dividing such interval into smaller intervals, and iteratively build the evolution operator at different time step based on the evolution operator obtained from the previous time step. These operations are algorithmic, and the recently introduced quantum singular value transformation framework allows us to algebraically manipulate these operations in a simple manner. Now we proceed to describe our main frameworks, beginning with general assumption regarding input, and subsequently building upon high order Runge-Kutta method and forward Euler method to construct two quantum algorithms that approximate the desired evolution operators under corresponding input assumptions.

\section{Main Framework}
\label{sec: mainframework}
In Heisenberg's picture, the dynamics of a quantum system under time-dependent Hamiltonian is governed by the following equation:
\begin{align}
    \frac{\partial U(t)}{\partial t} = -i H(t) U(t)
    \label{eqn: schrodinger}
\end{align}
The problem of simulating Hamiltonian is to use a universal quantum computer to construct the unitary operator $U$ such that:
\begin{align}
    | U - \tau \exp(-i \int_0^t H(s) ds) | \leq \epsilon
\end{align}
where $|.|$ refers to the spectral norm of matrix and $\tau$ is so-called time ordering operator. In the time-independent case, the integral $ \exp(-i \int_0^t H(s) ds) $ becomes $\exp(-i H t)$. Without loss of generality, we assume that we are interested in the evolution up to $t$ where $0 \leq t \leq 1$. We consider the case where the time-dependent Hamiltonian is decomposed as:
\begin{align}
    H(t) = \sum_{i=1}^m \gamma_i(t) H_i
\end{align}
where $m$ is some known factor, each $H_i$ is a time-independent Hamiltonian and $\gamma_i(t)$ is a time-dependent coefficient that is a (computable) function of $t$ with norm $|\gamma_i(t)| \leq 1$ for $0 \leq t \leq 1$. Suppose further that each $H_i$ can be (efficiently) simulated, for example, by by sparse-access method \cite{berry2007efficient}, truncated Taylor series \cite{berry2015simulating}, quantum walk \cite{berry2015hamiltonian}, etc. It means that our framework (soon to be outlined) is applicable whenever we can efficiently simulate the Hamiltonian under decomposition.  \\

For now, we assume that there is a mechanism for implementing $\exp(-i H_i t)$ for all $i=1,2,..., m$ and that each $H_i$ has norm at most $1/2$. Let $\mathcal{T}_i$ denotes the complexity (such as gate, or query to each $H_i$) for simulating $\exp(-i H_i t)$ and $\mathcal{T}_{\max} = \max_i \{ \mathcal{T}_i \}$. We use the following result of \cite{gilyen2019quantum}:
\begin{lemma}
    Suppose that $U = exp(-iH)$, where $H$ is a Hamiltonian of (spectral) norm $||H|| \leq 1/2$. Let $\epsilon \in (0,1/2]$ then we can implement a $(2/\pi, 2,\epsilon)$-block encoding of $H$ with $\mathcal{O}( \log(1/\epsilon)$ uses of controlled-U and its inverse, using $\mathcal{O}(\log(1/\epsilon)$ two-qubits gates and using a single ancilla qubit. 
\end{lemma}
It is simple to note that a unitary block encode itself. The unitary $\exp(-i H_i)$ (we opt $t=1$ in the simulation of $\exp(-i H_i t)$) can first be used to construct its controlled version (which is a standard result \cite{nielsen2002quantum}) and the above result allows us to use construct the block encoding of operator $\pi H_i/ 2$, for all $i$. Given that norm $||H_i|| \leq 1/2$, the factor $\pi/2$ can be removed using amplification technique \ref{lemma: amplification}, with further $\mathcal{O}(1)$ of such block encoding of $\pi H_i/2$. In particular, if we have oracle access to each $H_i$ (in a similar manner to \cite{berry2015simulating}), then without the simulation of $\exp(-i H_i)$, we can obtain the block encoding of $H_i$ directly for all $i$, if the norm $||H_i|| \leq 1/2$, according to lemma 48 of \cite{gilyen2019quantum}.

Now we proceed to outline the main quantum algorithm for simulating time-dependent Hamiltonian. At any time $0 \leq t \leq 1$, we use a single ancilla qubit initalized in $\ket{0}$ plus rotation gate to perform the following:
\begin{align}
    \ket{0} \longrightarrow t \ket{0} + \sqrt{1-t^2} \ket{1}
\end{align}
Then we can use the result in \cite{rattew2023non} (see also lemma \ref{lemma: diagonal} in appendix) to construct the block encoding of the operator 
\begin{align}
     \begin{pmatrix}
        t & 0 \\
        0 & \sqrt{1- t^2}
    \end{pmatrix}
\end{align}
Then, we can use quantum singular transformation framework \cite{gilyen2019quantum} to transform the above operator to:
\begin{align}
\label{eqn: 7}
     \begin{pmatrix}
        t & 0 \\
        0 & \sqrt{1- t^2}
    \end{pmatrix} \longrightarrow 
    \begin{pmatrix}
        \gamma_i(t) & 0 \\
        0 & \sqrt{1- \gamma_i(t)^2}
    \end{pmatrix}
\end{align}
for $i= 1,2,..,m$. Then we can use lemma \ref{lemma: tensorproduct} (a result of \cite{camps2020approximate}) to construct the block encoding of:
\begin{align}
    \begin{pmatrix}
        \gamma_i(t) & 0 \\
        0 & \sqrt{1- \gamma_i(t)^2}
    \end{pmatrix} \otimes H_i = \begin{pmatrix}
        \gamma_i(t) H_i & 0 \\
        0 & \sqrt{1- \gamma_i(t)^2} H_i
    \end{pmatrix}
\end{align}
for all $i$. Then we use lemma \ref{lemma: sumencoding} to construct the block encoding of:
\begin{align}
\label{eqn: Mi(t)}
    \frac{1}{m} \sum_{i=1}^m \begin{pmatrix}
        \gamma_i(t) H_i & 0 \\
        0 & \sqrt{1- \gamma_i(t)^2} H_i
    \end{pmatrix} = \frac{1}{m} \begin{pmatrix}
       \sum_{i=1}^m \gamma_i(t) H_i & 0 \\
        0 & \sum_{i=1}^m \sqrt{1- \gamma_i(t)^2} H_i
    \end{pmatrix} \equiv M_i(t)
\end{align}
If, for instance, we know that the norm $||H||$ is less than $1/2$, the factor $m$ from the above equation can be removed using amplification technique \cite{gilyen2019quantum} (see also lemma \ref{lemma: amplification} in appendix) with further $\mathcal{O}(m)$ complexity, i.e., with $m$ usage of such $M_i(t)$. We further note that, after we perform amplification, the above operator is actually block encoding $ \sum_{i=1}^m \gamma_i(t) H_i \equiv H(t)$.  Therefore, the above procedure allows us to construct a block encoding of $H(t)$, for any given $t$, up to accuracy $\epsilon$. The circuit complexity of this step is $\mathcal{O}(m^2 \mathcal{T}_{max} \log(\frac{1}{\epsilon}))$ where $\mathcal{T}_{max} = \max_{i} \{ \mathcal{T}_i \}$, as we first need to use lemma \ref{lemma: sumencoding} to construct the (uniform) linear combination, then we use amplification to remove the factor $m$.  \\

\subsection{Runge-Kutta Method}
\label{sec: rungekutta} 

Now suppose that at $t=0$, we have $U(t=0) = \Ibb$. Recall that the dynamics of unitary is:
\begin{align}
    \frac{\partial U(t)}{\partial t} = -i H(t) U(t)
\end{align}
We break the time interval $[0,1]$ into discrete steps as $[0, \Delta t, 2\Delta t, 3\Delta t, ..., N\Delta t \equiv 1]$, and we are aiming at finding the unitary at corresponding steps. At the $n$-th time step, the $p$-th order Runge-Kutta method computes the unitary $U( (n+1)\Delta t)$ of the next time step as following:
\begin{align}
    U((n+1)\Delta t) = U(n\Delta t) + \Delta t \sum_{j=1}^s b_j k_j \\
    k_j = -i H( n\Delta t+ c_j \Delta t ) \Big( U(n\Delta t) + \Delta t \sum_{m=1}^{j-1} a_{jm} k_m \Big)
\end{align}
Given that there exists a (classical) procedure for computing all coefficients $\{ a_{jm}, b_j,c_j \}$, lemma \ref{lemma: sumencoding} and \ref{lemma: product} allows us to construct the block encoding of $U( (n+1)\Delta t)$. More concretely, the block encoding of $H( n\Delta t+ c_j \Delta t )$ can be constructed via the construction of $M_i(t)$ above, i.e., by setting $t = n\Delta + c_j \Delta t $. Then, beginning with $k_1 = -i H(n\Delta t) U(n\Delta t)$, which can be constructed by using lemma \ref{lemma: product} with operators $H(n\Delta t), U(n\Delta t) $, then we have that:
\begin{align}
    k_2 &= -i H(n\Delta t + c_2 \Delta t) \Big( U(n\Delta t) + \Delta t a_{21} k_1  \Big) \\
    &= -i H(n\Delta t + c_2 \Delta t) \Big( U(n\Delta t) + \Delta t a_{21} (-i H(n\Delta t) U(n\Delta t))  \Big)
\end{align}
Similarly:
\begin{align}
    k_3 = -i H(n\Delta t + c_3 \Delta t) \Big(  U(n\Delta t) + \Delta t a_{31} k_1 + \Delta t a_{32} k_2   \Big) 
\end{align}
and higher orders are proceeded in a straightforward manner. To construct a block encoding of $k_2$, we first use lemma \ref{lemma: scale} with a scaling factor $\Delta t a_{21}$ to construct the block encoding of $\Delta t a_{21} k_1$. Then we use lemma \ref{lemma: sumencoding} to construct the block encoding of $\frac{1}{2}( U(n\Delta t) + \Delta t a_{21} k_1  )$, followed by using lemma \ref{lemma: product} to construct the block encoding of $- \frac{i}{2} H(n\Delta t + c_2 \Delta t) \Big( U (n\Delta t) + \Delta t a_{21} k_1)  = k_2/2$. From the block encoding of $k_1, k_2/2$, we can construct the block encoding of $k_3/3$ similarly by using lemma \ref{lemma: sumencoding} to first construct the block encoding of $\frac{1}{3}(  U(n\Delta t) + \Delta t a_{31} k_1 + 2 \Delta t a_{32} \frac{k_2}{2}) = \frac{k_3}{3}$. Proceeding similarly, we can obtain a block encoding of $k_j/j$ for remaining $j$. From such block encodings, we again use lemma \ref{lemma: scale} to insert the scaling factor $\Delta t jb_j$ to each operator $k_j/j$. Then we use \ref{lemma: sumencoding} to construct the block encoding of $\frac{\Delta t}{s} \sum_{j=1}^s j b_j k_j$. From such block encoding, we use lemma \ref{lemma: sumencoding} to construct the block encoding of $\frac{1}{2s}(  U(n\Delta t) + \Delta t \sum_{j=1}^s b_j k_j ) = \frac{1}{2s} U( (n+1)\Delta t)$. The factor $s$ can be removed by using amplification method, e.g., lemma \ref{lemma: amplification}, with $\mathcal{O}(s)$ use of block encoding of $U( (n+1)\Delta t)$.  In an exact manner, one can continue and construct $\frac{1}{2} U( (n+2)\Delta t)$ from $\frac{1}{2} U( (n+1)\Delta t)$, and so on, thus completing the method for obtaining the unitaries $[ U(0), U(\Delta t), U(2\Delta t), ..., U(N\Delta t) ]$ up to a factor $\frac{1}{2}$. \\

To analyze the quantum circuit depth, we observe that at the $n$-th time step, the algorithm takes $\mathcal{O}(m^2 \mathcal{T}_{max} \log(\frac{1}{\epsilon}))$ to construct the block encoding of $H(n\Delta t + c_j \Delta t)$ (up to accuracy $\epsilon$)  for all $j$. Then it takes $\mathcal{O}(j)$ to construct $k_j/j$. Hence, to construct $\Delta t \sum_{j=1}^s b_j k_j$, the circuit depth would accumulate as $\mathcal{O}( \sum_{j=1}^s j) = \mathcal{O}(s^2)$. Eventually, we need to use amplification to remove the factor $s$, resulting further $\mathcal{O}(s)$ usage of block encoding of $\frac{1}{2s} U( (n+1)\Delta t)$. Hence, if we denote $T_n$ as the circuit depth required for constructing the block encoding of $\frac{1}{2} U(n\Delta t)$. Then the total circuit depth to go from $ \frac{1}{2} U(n\Delta t)$ to $\frac{1}{2} U( (n+1)\Delta t)$ is then $\mathcal{O}( (m^2 + sT_n + s^2 ) s  )$. By the same procedure, $T_n = \mathcal{O}(  (m^2 \mathcal{T}_{max} \log(\frac{1}{\epsilon})+ sT_{n-1} +s^2)s) $. Proceeding inductively, one can deduce that the total circuit depth for constructing $\frac{1}{2} U(N\Delta t) = \mathcal{O}( (m^2 \mathcal{T}_{max} \log(\frac{1}{\epsilon})s+ s^3) s^N )$. 

As analyzed in the standard text \cite{butcher1996history}, the error induced in this $p$-th order Runge-Kutta method follows as $\mathcal{O}( \Delta t^p  )$. Hence, if we wish the error to be $\epsilon$, then we set $\Delta t^p = \epsilon$. Replacing $\Delta t = t/N$ leading to:
\begin{align}
    \frac{t}{ N} =\epsilon^{1/p} \longrightarrow N = \frac{t}{\epsilon^{1/p}}
\end{align}
Then the overall quantum circuit depth is 
$$ \mathcal{O}( (m^2 \mathcal{T}_{max} \log(\frac{1}{\epsilon})s+s^3) s^{t/(\epsilon^{1/p})} )$$
We further mention that for a Runge-Kutta of order $p$, the value of $s$ will be $\mathcal{O}(p)$. We remind that $\mathcal{T}_{max} = \max \{ \mathcal{T}_i \}_{i=1}^m$ where each $\mathcal{T}_i$ is the complexity for simulating individual term $H_i$. In appropriate setting, optimal time-independent Hamiltonian result has been found \cite{low2017optimal, berry2015hamiltonian}. The complexity $\mathcal{T}_i$ of such simulation of $\exp(-i H_i)$ grows as $\mathcal{O}( ||H_i|| d_i + \log( \frac{1}{\epsilon})  )$ where $d_i$ is the sparsity of $H_i$. Hence, $\mathcal{T}_{\max} = \mathcal{O}( d_{\max} ||H||_{\max} + \log (1/\epsilon) )$.

We summarize our result in the following theorem.
\begin{theorem}[High-Order Runge-Kutta]
    Let the Hamiltonian $H = \sum_{i=1}^m \gamma_i(t) H_i$ where $H_i$ is time-independent Hamiltonian. Suppose that $|\gamma_i(t)| \leq 1$ for $0 < t < 1$ and the norm $||H||, ||H_i|| \leq 1/2$ for all $i$. Suppose further that each $H_i$ having sparsity $d_i$ can be efficiently simulated. Let $d_{\max} = \max_i \{ d_i\}$ and $||H||_{\max} = \max_i \{||H_i|| \}$. Then the evolution operator $U(t)$ can be simulated up to additive accuracy $\epsilon$ in complexity 
    $$ \mathcal{O}( (m^2 (d_{\max} ||H||_{\max} + \log \frac{1}{\epsilon} ) \log(\frac{1}{\epsilon})p+p^3) p^{t/(\epsilon^{1/p})} )$$
    where $p$ is arbitrary integer. 
\end{theorem}
The above complexity matches the optimal scaling (up to some factor) on norm $||H||$ and sparsity parameter $d$. The dependence on time $t$ and inverse of error $1/\epsilon$ is exponential, which is probably far from optimal, as established in time-independent setting \cite{berry2015simulating} to be linear in $t$. 

\subsection{High Order Forward Euler}
\label{sec: forwardeuler}
Runge-Kutta method works by evaluating the right-hand side of equation \ref{eqn: schrodinger} at multiple points. Instead, forward Euler method works by evaluating the first derivative at given point, and generalization to higher order pretty much mimic Taylor expansion. More specifically, at a given $n$-th time step (as defined above for Runge-Kutta), the next unitary $U((n+1)\Delta t)$ is updated as:
\begin{align}
    U((n+1)\Delta t) \approx U( n\Delta t) + \sum_{j=1}^p \frac{1}{j!}\frac{\partial^j U((n+1)\Delta t)}{\partial^j t} (\Delta t)^j
\end{align}
From the Schrodinger's equation:
\begin{align}
    \frac{\partial U(t)}{\partial t} = -i H(t) U(t)
\end{align}
Then the second-order derivative is:
\begin{align}
    \frac{\partial^2 U(t)}{\partial^2 t} &= - i \frac{\partial (H(t)U(t))}{\partial t} \\
    &= -i \Big(  \frac{\partial H(t)}{\partial t}U(t) + H(t) \frac{\partial U(t)}{\partial t} \Big)\\
    &= -i \Big((  \frac{\partial H(t)}{\partial t} + H(t) (-i H(t) )  ) U(t)    \Big) \\
    &= -i \Big(   ( \frac{\partial H(t)}{\partial t} - i H^2(t)  ) U(t) \Big) \\
\end{align}
Taking further derivative, we have that:
\begin{align}
    \frac{\partial^3 U(t)}{\partial^3 t} &= -i (  \frac{\partial^2 H(t)}{\partial t} + \frac{\partial H(t)}{\partial t}  ) (-iH(t) )U(t) +  (-i)( \frac{\partial H(t)}{\partial t}U(t) + H(t) ) \\
    &= -i \Big( ( \frac{\partial^2 H(t)}{\partial^2 t} - 3i \frac{H(t) \partial H(t)}{\partial t} - H^3(t)          ) U(t) \Big)
\end{align}
We remark a crucial point, that is the higher order derivative of $U$ with respect to time $t$ can be decomposed products of two operators. The first one is only consisted of Hamiltonian and its high derivatives with respect to time $t$, meanwhile the second operator is $U(t)$. Therefore, for a $p$-th order Taylor series, we have:
\begin{align}
    U( (n+1) \Delta t) &\approx U(n\Delta t) + \sum_{j=1}^p \frac{1}{j!} f_j(H) (\Delta t)^j U(n\Delta t) \\
    &= \Big( I +  \sum_{j=1}^p \frac{1}{j!} f_j(H) (\Delta t)^j   \Big) U(n\Delta t) 
\end{align}
where $f_j(H) = \frac{\partial^j U(t)}{\partial^j t}$ contains power of $H$ plus its (higher) derivatives with respect to $t$, as we have provided some examples above for second and third-order derivation. We have assumed that:
\begin{align}
    H(t) = \sum_{i=1}^m \gamma_i(t) H_i
\end{align}
Therefore, arbitrary high order derivative of $H(t)$ is:
\begin{align}
    \frac{\partial^j H(t)}{\partial^j t } =  \sum_{i=1}^m \frac{\partial^j \gamma_i(t)}{\partial^j t} H_i
\end{align}
Given that each $\gamma_i(t)$ is a computable function of $t$, its derivative is also computable. Let $M_j$ the the upper bound of $j$-th order derivative of $H$ with respect to $t$, i.e.
\begin{align}
    M_j = \max_{t \in [0,1]} ||\frac{\partial^j H(t)}{\partial^j t }||
\end{align}
Using a similar procedure for obtaining $M_i(t) $ (see equation \ref{eqn: Mi(t)}), for example, in equation \ref{eqn: 7}, we can choose the corresponding derivative $\frac{1}{M_j} \frac{\partial^j \gamma_i(t)}{\partial^j t}$ instead. Hence, one can construct the block encoding of $ \frac{1}{M_j} \frac{\partial^j H(t)}{\partial^j t }$ for all $j$. Let $M = \max_j \{M_j \}$. We can use lemma \ref{lemma: scale} to transform the block encoding of $\frac{1}{M_j} \frac{\partial^j H(t)}{\partial^j t }$ to the block encoding of $ \frac{1}{M} \frac{\partial^j H(t)}{\partial^j t }$ by using scaling factor $M_j/M$ in lemma \ref{lemma: scale}. Then we proceed to use lemma \ref{lemma: sumencoding} to construct the block encoding of $f_j(H)/(jM)$. The factor $j$ comes from the number of terms in the corresponding derivative of $U$ with respect to $t$. For example, in the second and third  derivative 
\begin{align}
    \frac{\partial^2 U(t)}{\partial^2 t} &= -i \Big(   ( \frac{\partial H(t)}{\partial t} - i H^2(t)  ) U(t) \Big) \\
    \frac{\partial^3 U(t)}{\partial^3 t} &= -i \Big( ( \frac{\partial^2 H(t)}{\partial^2 t} - 3i \frac{H(t) \partial H(t)}{\partial t} - H^3(t)          ) U(t) \Big) \\
\end{align}
The operator besides $U(t)$ has total number of $2$ and $3$ terms, respectively. Hence, using lemma \ref{lemma: sumencoding} to form the (block encoding of) uniform linear combination of those terms would incur the factor $1/2$ and $1/3$ for the above two equations. For higher $j$-th order derivative, the factor is then $1/j$. Then, we can use lemma \ref{lemma: scale} to multiply the block encoding of $f_j(H)/(jM)$ with a factor $f_j(H)/(jM) \cdot (\Delta t)^j / (j-1)! = (f_j(H) \Delta t^j)/(M j!) $. Then we use lemma \ref{lemma: sumencoding} again to construct the block encoding of operator:
$$ \frac{1}{M(p+1)} ( I  + \sum_{j=1}^p \frac{1}{j!} f_j(H) (\Delta t)^j )   $$
Then we use lemma \ref{lemma: product} to construct the block encoding of 
$$ \frac{1}{M(p+1)} ( I  + \sum_{j=1}^p \frac{1}{j!} f_j(H) (\Delta t)^j ) U(n\Delta t)  $$
The factor $M(p+1)$ can be removed using amplification technique \ref{lemma: amplification}, resulting a further $\mathcal{O}(Mp)$ usage of block encoding of the above operator. 

To analyze the complexity, we revise the method as follows. At $n$-th time step, we need to construct the $p$ block encodings of operator $(1/Mj!) f_j(H) (\Delta t)^j$ for $j=1,2,...,p$. Each block encoding of $f_j(H)$ is constructed from the block encoding of $H(n\Delta t)$ and its higher derivatives, having $j$ terms totally. The complexity required to construct $H(n\Delta t)$ as described previously is $\mathcal{O}(m^2 \mathcal{T}_{\max}\log(\frac{1}{\epsilon})  )$ where $\mathcal{T}_{\max}$ is defined in previous section as $\mathcal{T}_{\max} = \max_i \{\mathcal{T}_i \}$ where $\mathcal{T}_i$ is the complexity required to simulate $H_i$ (for a unit time) in the decomposition $H(t) = \sum_{i=1}^m \gamma_i(t) H_i$. As $H_i$ is time-independent, we have assumed that $H_i$ could be simulated using standard approaches \cite{low2017optimal, berry2007efficient, berry2015hamiltonian}. Thus, the best known complexity is $\mathcal{O}( d_i ||H_i||  + \log(\frac{1}{\epsilon})  )$ where $d_i$ is the sparsity of $H_i$. Therefore, the circuit depth required to construct the block encoding of $f_j(H)$ is $\mathcal{O}( j m^2 \mathcal{T}_{\max} \log(\frac{1}{\epsilon}) ) = \mathcal{O}(j m^2 (d_{\max} ||H||_{\max}  + \log \frac{1}{\epsilon}) \log(\frac{1}{\epsilon})   ) $. To construct the block encoding of
$$ \frac{1}{(p+1)} ( I  + \sum_{j=1}^p \frac{1}{j!} f_j(H) (\Delta t)^j )  $$
which uses the block encoding of $f_j(H)$ each once, then the total circuit depth resulted in is:
\begin{align}
    \mathcal{O}(  \sum_{j=1}^p j m^2 (d_{\max} ||H||_{\max}  + \log \frac{1}{\epsilon})\log(\frac{1}{\epsilon})  ) = \mathcal{O}( p^2 m^2 (d_{\max} ||H||_{\max}  + \log \frac{1}{\epsilon})\log(\frac{1}{\epsilon})  )
\end{align}
The last step is using lemma \ref{lemma: product} to construct the block encoding of
$$ \frac{1}{M(p+1)} ( I  + \sum_{j=1}^p \frac{1}{j!} f_j(H) (\Delta t)^j ) U(n\Delta t)  $$
then use amplification technique (lemma \ref{lemma: amplification}) to remove the factor $M(p+1)$. The final accumulated circuit depth is $\mathcal{O}(M p^3 m^2 (d_{\max} ||H||_{\max}  + \log \frac{1}{\epsilon}) ) $. We remind that this is the circuit depth required to go from $U(n\Delta t)$ to $U( (n+1)\Delta t)$. According to standard Taylor approximation analysis, the overall error of this method is $\mathcal{O}( \Delta t^p) $. If we set the overall error to $\epsilon$, then we have that $\Delta t \in \mathcal{O} (\epsilon^{1/p})$, which means that $t/N \in \mathcal{O}(\epsilon^{1/p}) $ and hence, $N \in \mathcal{O}( t/\epsilon^{1/p})$, which is also the total number of iterations we need to perform. We summarize the result of high order Taylor series in the following theorem.
\begin{theorem}[High-Order Taylor Series]
    Let the Hamiltonian $H = \sum_{i=1}^m \gamma_i(t) H_i$ where $H_i$ is time-independent Hamiltonian. Suppose that $|\gamma_i(t)| \leq 1$ for $0 < t < 1$ and higher derivative of all $\gamma_i(t)$ is bounded, computable and the norm $||H||, ||H_i|| \leq 1/2$ for all $i$. Suppose further that each $H_i$ having sparsity $d_i$ can be efficiently simulated. Let 
    $$d_{\max} = \max_i \{d_i\}, ||H||_{\max} = \max_i \{  || H_i \} \text{ and } M = \max_j \max_{t \in [0,1]} || \frac{\partial^j H}{\partial t^j}  ||$$. 
    Then the evolution operator $U(t)$ can be simulated up to additive accuracy $\epsilon$ in complexity 
    $$ \mathcal{O}( M p^3 d_{\max} m^2 ( d_{\max} ||H||_{\max}  + \log \frac{1}{\epsilon}) \frac{t}{\epsilon^{1/p}} \log(\frac{1}{\epsilon})  )$$
    where $p$ is arbitrary integer. 
\end{theorem}

Comparing between this approach and previous approach, we can see that Taylor series achieve better scaling on most parameters, especially on inverse of error tolerance and time. The scaling on time $t$ is optimal, as it was proved before \cite{berry2015hamiltonian, childs2017lecture} that it is impossible to simulate Hamiltonian sublinearly in time $t$. 

\section{Discussion and Conclusion}
\label{sec: discussandconclusion}
In this work, we have provided two newly alternative quantum algorithms for simulating time-dependent Hamiltonian. Our approach is inspired by classical numerical method, such as Runge-Kutta and forward Euler method. The input Hamiltonian model of our work is also different from previous works, such as \cite{berry2020time, kieferova2019simulating}, meanwhile sharing a bit of similarity to \cite{an2022time}. Under this model, our Runge-Kutta based method achieves exponential time dependence on time $t$, meanwhile having linear dependence on sparsity as well as Hamiltonian norm, which is probably optimal. The second approach relies on high order forward Euler method perform better than Runge-Kutta method overall, achieving linear dependence on time $t$ as well as sparsity and norm, which is arguably optimal in these parameters. 

Our examples have offered a new way to simulate time-dependent Hamiltonian, which could be understood as quantizing numerical method for solving Schrodinger equation. Thus, it is highly desired to expand it further to different techniques, making further uses of quantum resources to probe quantum system. For example, in \cite{peskin1994solution}, the authors introduced the propagators method for solving time-dependent Schrodinger equation, with specific application in atomic-laser interaction. Quantum singular value transformation framework was shown to be able to involve Chebyshev series in a simple way, thereby, it is very motivating to explore how it can be uses to simulate, or understand the dynamics of time-dependent Hamiltonian further using quantum techniques. We leave these questions for future exploration.

\subsection*{Acknowledgement}
The author thanks Trung V. Phan for interesting discussion in related projects. We acknowledge support from Center for Distributed Quantum Processing, Stony Brook University

\bibliography{ref.bib}{}
\bibliographystyle{unsrt}

\clearpage
\newpage
\onecolumngrid
\appendix

\begin{definition}
\label{def: blockencode}
    Suppose that $A$ is an s-qubit operator, $\alpha, \epsilon \in \mathbb{R}_+$ and $a \in \mathbb{N}$, then we say that the $(s+a)$-qubit unitary $U$ is an $(\alpha, a ,\epsilon)$-block encoding of $A$, if
    $$ || A - \alpha (\bra{0}^{\otimes a} \otimes \mathbb{I}) U ( \ket{0}^{\otimes a} \otimes \mathbb{I} ) || \leq \epsilon$$
    Equivalently, in matrix representation, $U$ is said to be a block encoding of $A/\alpha$ if $U$ has the form
    \begin{align*}
    U = \begin{pmatrix}
       \frac{A}{\alpha} & \cdot \\
       \cdot & \cdot \\
    \end{pmatrix}.
\end{align*}
where $(.)$ in the above matrix refers to irrelevant blocks that could be non-zero.
\end{definition}

\begin{lemma}[Theorem 2 in \cite{rattew2023non}]
\label{lemma: diagonal}
     Given an n-qubit quantum state specified by a state-preparation-unitary $U$, such that $\ket{\psi}_n=U\ket{0}_n=\sum^{N-1}_{k=0}\psi_k \ket{k}_n$ (with $\psi_k \in \mathbb{C}$ and $N=2^n$), we can prepare an exact block-encoding $U_A$ of the diagonal matrix $A = {\rm diag}(\psi_0, ...,\psi_{N-1})$ with $\mathcal{O}(n)$ circuit depth and a total of $\mathcal{O}(1)$ queries to a controlled-$U$ gate  with $n+3$ ancillary qubits.
\end{lemma}

\begin{lemma}[\cite{camps2020approximate}]
\label{lemma: tensorproduct}
    Given the unitary block encoding $\{U_i\}_{i=1}^m$ of multiple operators $\{M_i\}_{i=1}^m$ (assumed to be exact encoding), then, there is a procedure that produces the unitary block encoding operator of $\bigotimes_{i=1}^m M_i$, which requires a single use of each $\{U_i\}_{i=1}^m$ and $\mathcal{O}(1)$ SWAP gates. 
\end{lemma}

\begin{lemma}[Linear combination of block-encoded matrices]
    Given unitary block encoding of multiple operators $\{M_i\}_{i=1}^m$ and a unitary that prepares the state $\ket{y} = \sum_{i=1}^{m} \sqrt{ sign(y_i) y_i/ \beta } \ket{i}$ where $\beta = \sum_{i=1}^{m} |y_i|$. Then, there is a procedure that produces a unitary block encoding operator of \,$\sum_{i=1}^m  y_i M_i/ \beta $ in complexity $\mathcal{O}(m)$, using block encoding of each operator $M_i$ a single time. 
    \label{lemma: sumencoding}
\end{lemma}

\begin{lemma}[Block Encoding of Product of Two Matrices]
\label{lemma: product}
    Given the unitary block encoding of two matrices $A_1$ and $A_2$ (assuming to have norm less than 1), then there exists an efficient procedure that constructs a unitary block encoding of $A_1 A_2$ using each block encoding of $A_1,A_2$ one time. 
\end{lemma}

\begin{lemma}\label{lemma: amp_amp}[\cite{gilyen2019quantum} Theorem 30]
\label{lemma: amplification}
Let $U$, $\Pi$, $\widetilde{\Pi} \in {\rm End}(\mathcal{H}_U)$ be linear operators on $\mathcal{H}_U$ such that $U$ is a unitary, and $\Pi$, $\widetilde{\Pi}$ are orthogonal projectors. 
Let $\gamma>1$ and $\delta,\epsilon \in (0,\frac{1}{2})$. 
Suppose that $\widetilde{\Pi}U\Pi=W \Sigma V^\dagger=\sum_{i}\varsigma_i\ket{w_i}\bra{v_i}$ is a singular value decomposition. 
Then there is an $m= \mathcal{O} \Big(\frac{\gamma}{\delta}
\log \left(\frac{\gamma}{\epsilon} \right)\Big)$ and an efficiently computable $\Phi\in\mathbb{R}^m$ such that
\begin{equation}
\left(\bra{+}\otimes\widetilde{\Pi}_{\leq\frac{1-\delta}{\gamma}}\right)U_\Phi \left(\ket{+}\otimes\Pi_{\leq\frac{1-\delta}{\gamma}}\right)=\sum_{i\colon\varsigma_i\leq \frac{1-\delta}{\gamma} }\tilde{\varsigma}_i\ket{w_i}\bra{v_i} , \text{ where } \Big|\!\Big|\frac{\tilde{\varsigma}_i}{\gamma\varsigma_i}-1 \Big|\!\Big|\leq \epsilon.
\end{equation}
Moreover, $U_\Phi$ can be implemented using a single ancilla qubit with $m$ uses of $U$ and $U^\dagger$, $m$ uses of C$_\Pi$NOT and $m$ uses of C$_{\widetilde{\Pi}}$NOT gates and $m$ single qubit gates.
Here,
\begin{itemize}
\item C$_\Pi$NOT$:=X \otimes \Pi + I \otimes (I - \Pi)$ and a similar definition for C$_{\widetilde{\Pi}}$NOT; see Definition 2 in \cite{gilyen2019quantum},
\item $U_\Phi$: alternating phase modulation sequence; see Definition 15 in \cite{gilyen2019quantum},
\item $\Pi_{\leq \delta}$, $\widetilde{\Pi}_{\leq \delta}$: singular value threshold projectors; see Definition 24 in \cite{gilyen2019quantum}.
\end{itemize}
\end{lemma}
\begin{lemma}[Scaling Block encoding] 
\label{lemma: scale}
    Given a block encoding of some matrix $A$ (as in~\ref{def: blockencode}), then the block encoding of $A/p$ where $p > 1$ can be prepared with an extra $\mathcal{O}(1)$ cost.  
\end{lemma}

\end{document}